%Paper: hep-ph/9303268
%From: DATWOOD@SLACVM.SLAC.Stanford.EDU
%Date: Tue, 16 Mar 1993   14:16 -0800 (PST)

%
%
%  Final version of polarization paper.
%  this is version submitted to prl on 3/12/93
%
%  Postscript version of figure 1 is immediately after \end command
%
%
%
%
%
%
%
%
%
%
%
%
%%%%%%%%%%%%%%%%%%%%%%%%%%%%%%%%%%%%%%%%%%%%%%%%%%%%%%%%%%%%%%%%%%%%%%%%%%%%%%
%
%
%    set up mode draft=0 is draft mode  draft=1 is final mode
%
\newcount\draft
\newcount \titlepg
\newcount  \intro
\global\draft=        1
\global\titlepg=      0
\global\intro=      0
%
%
%
%
%
%
%    definitions
%
%

%

%
\def\uglu{\hskip 0pt plus 1fil minus 1fil}
\def\uglux{\hskip 0pt plus .75fil minus .75fil}
\def\slashed#1{\setbox200=\hbox{$ #1 $}
    \hbox{\box200 \hskip -\wd200 \hbox to \wd200 {\uglu $/$ \uglux}}}

\def\slp{\slashed p}

\def\g5{\gamma_5}

%
%
%

%
%
%           SLAC preprint header
%
\newcount\aaa
\newcount\aab
\newcount\aac
\newcount\aad
\global\aaa=1
\global\aab=1
\global\aac=1
\global\aad=1
\def\lastx{\the\aaa}
\def\lasty{\the\aab}
\def\lastz{\the\aac}
\def\ref#1{$^#1$}
\def\xnum#1{\the\aaa
\ifcase\draft\string#1\fi
\xdef#1{\the\aaa\ifcase\draft\string#1\fi}
\xdef\lastx{\the\aaa}
\global\advance\aaa by 1}
\def\eqnox#1{\eqno{(\xnum{#1})}}
\def\ynum#1{\the\aab\xdef#1
{\the\aab}\xdef\lasty{\the\aab}
\global\advance\aab by 1}
\def\znum#1{\the\aac\xdef#1
{\the\aac}\xdef\lastz{\the\aac}
\global\advance\aac by 1}
\def\refnum#1{\ref{\the\aad\ifcase\draft\string#1\fi}
\xdef#1{\the\aad\ifcase\draft\string#1\fi}
\xdef\lastref{\the\aad}
\global\advance\aad by 1}
\def\refset#1{
\xdef#1{\the\aad\ifcase\draft\string#1\fi}
\xdef\lastref{\the\aad}
\global\advance\aad by 1}
\def\sqref#1{[{\the\aad\ifcase\draft\string#1\fi}]
\xdef#1{\the\aad\ifcase\draft\string#1\fi}\xdef\lastref{\the\aad}
\global\advance\aad by 1}
\footline={\hfil}
\def\deletion#1{\centerline{---deletion---}}
\def\ref#1{${}^{\scriptstyle#1\atop {\ }}$}

\magnification=\magstep1
%
%     date printing routine
%
\newcount\aba
\newcount\abb
\newcount\abc
\newcount\abd
\def\dater{
\aba=\time
\abb=\time
\divide\aba by 60
\abd=\aba
\multiply\abd by -60
\advance\abb by \abd
\ifnum\abb<10
            {\the\aba:0\the\abb/\the\day/\the\month/\the\year}\fi
\ifnum\abb>9
            { \the\aba:\the\abb/\the\day/\the\month/\the\year}\fi
}
\def\themonth{
\ifcase\month\or January
              \or February
               \or March
                \or April
                 \or May
                  \or June
                   \or July
                    \or August
                     \or September
                      \or October
                       \or November
                        \or December
                         \or Never
\fi}
%
%
%
%
%     some font definitions
%
\font\fta=cmr10 scaled\magstep1
\font\ftb=cmbx10 scaled\magstep1

%
%                 ##
%\ifcase\%  \iffalse
%
%
%     banner
%
{\fta\hfil Stanford Linear Accelerator Center \hfil}
\medskip
\medskip
\hrule
\vskip 1pt
\hrule
\smallskip
{\themonth \the\year\hfill SLAC-PUB-6083
}
\smallskip
\hrule
\vskip 1pt
\hrule
\medskip
%
%
%     #date/time stamp
%
%
\vfill\vfill
%
%     #title
%
\centerline{\ftb
Transverse Tau Polarization in Decays of the Top and Bottom
Quarks
}
\centerline{\ftb
in the Weinberg Model of CP Non-conservation
}
\vfill
%
%
%     #authors
%
{\baselineskip=12pt
\centerline {
D. Atwood*, G. Eilam\dag\   and A. Soni\ddag}
\bigskip
\bigskip
\bigskip
{\obeylines
*       Dept. of Physics, SLAC, Box 4349 Stanford CA 94309 USA.
\dag\    Dept. of Physics, Technion, Haifa, Israel
\ddag\   Dept. of Physics, Brookhaven National Laboratory, Upton NY 11973 USA.
}}
\vfill
%
%
%
%     #abstract
%
{\baselineskip=12pt
\centerline{\bf ABSTRACT}
\bigskip
We show that the transverse polarization asymmetry of the
$\tau$-lepton in the
decay $t\rightarrow b \tau \nu$ is extremely sensitive to CP
violating phases arising from the charged Higgs exchange in the Weinberg
model of CP non-conservation. Qualitatively, the polarization asymmetries
are
enhanced over rate or energy asymmetries by a factor of $\approx
{m_t\over m_\tau}\approx
O(100)$. Thus for optimal values of the parameters
the method requires $\approx 10^4$ top pairs to be observable rather
than $10^7$ needed for rate or energy asymmetries.
We also examine $\tau$ polarization in b decays via $b \rightarrow
c \nu \tau$ and find that it can also be very effective
in constraining the CP violation parameters of the
extended Higgs sector.
}
\vfill
\vfill
\eject
%
%
%  start of text proper  #####
%
\footline={\hss\tenrm\folio\hss}\pageno=1
\baselineskip=17.5pt
%-------------============########## \baselineskip=27pt
%
%
%\hbox{\bf Introduction and Motivation}
%
%
%
Due to its mass scale the top quark
represents a unique probe for addressing
to the long-standing issue of CP violation.
While the FNAL Tevatron is expected to produce
enough top quarks to establish its existence, the hadronic and
$e^+-e^-$ colliders under construction and being proposed
should facilitate studies related to CP. In this context,
the search for optimal experimental strategies is clearly
important and requires extended phenomenological studies.
\refset{\ehs}
\refset{\km}
\refset{\nelson}
\refset{\aas}
\refset{\pra}
\refset{\jg}
\refset{\others}
\refset{\refone}
\refset{\rxu}
\refset{\mpcs}
\par
In the Standard Model (SM) CP violation effects in the top quark
are too small to be observable\ref\ehs. However, it is difficult to see
why the Kobayashi-Maskawa\ref\km
(KM) phase should be the only source for CP non-conservation.
Extensions of the SM almost invariably lead to additional
CP violating phases.
Indeed the observed baryon asymmetry in the
universe is often used to argue that additional sources of
CP beyond the KM phase are a necessity\ref\nelson. We are therefore motivated
to continue our investigation of CP non-conservation
caused in the production\ref\aas  and decays\ref{\pra,\jg}
of the top quark in one
popular extension of the SM\ref\others,
namely the Weinberg Model\ref\refone (WM).
This model leads to large, possibly observable, dipole moments\ref\rxu and
rather sizable partially integrated rate asymmetry\ref\pra (PIRA) in the
semi-leptonic modes $t \rightarrow b \tau \nu_{\tau}$. In
this work we show that the transverse polarization of the
$\tau$ in these semileptonic transitions is extremely
sensitive to CP violation effects due to the extended Higgs
sector.
Thereby, with optimal values of the parameters
in the WM, requiring only a few thousand top quark pairs rather than
$ \ge10^7$ needed for the rate, the triple
correlation or the energy asymmetries\ref{\pra,\jg,\mpcs}.
Thus the polarization effects
become accessible not only to SSC/LHC, where the
estimates are for $10^7 - 10^8$ top pairs/year but also
to an electron-positron linear collider with an anticipated rate
of about $10^4$/yr.
%%%%%%
%
%     In fact given the uncertainties in the parameters of the
%     underlying model, it may even be possible to
%     put useful constraints on the model
%     through measurements of the $\tau$
%     polarization at the Tevatron.
%
%%%%%%
Clearly this approach can only be used if the top
detectors are capable of measuring the $\tau$
polarization.

We have also examined the effectiveness
of $\tau$ polarization as a possible signal of
CP violation in
semi-leptonic b decays
due to the extended Higgs sector.
%In Figs 1 and 2 we then need to
%make the simple replacement of $t \rightarrow b$
%with $b \rightarrow c$.
Our finding is that
one of the transverse polarization
asymmetries is useful in b decays as well
as it can lead to stringent constraints
on the parameters of the Higgs model with
about $10^6$ B mesons.
\refset{\mhgemt}
\par
The semi-leptonic decay of the top quark proceeds through W exchange
and Higgs exchange graphs (See Fig. 1a and 1b). The CP
violating phase is in the Higgs exchange. Since $m_t >
m_W$, the dominant contribution of the W-graph is
from on shell W-bosons. Thus
Fig 1a possesses both dispersive and absorptive
pieces, and indeed both can have
significant resonance enhancement, as we shall see below
\ref{\ehs,\pra,\jg}. The absorptive
piece of Fig 1a interferes with the CP violating
Higgs phase to contribute to
observables (e.g. rate asymmetry) that are
CP-odd, $T_N$ even\ref{\mhgemt}.
Here $T_N$ denotes naive time reversal where the spins and momenta of all
particles are reversed but initial and final states are not interchanged.
On the other hand
observables such as momentum
triple correlations that are CP-odd,
$T_N$-odd receive contribution from
the dispersive part of Fig. 1a.
\par
Let us now qualitatively try to understand
why the transverse polarization is much more sensitive
compared to energy or rate asymmetries.
Recall that PIRA (or energy asymmetry) goes as\refnum\seepra
$m_{\tau}^2/m_H^2$.
It is important to understand the sources for
the two powers of $m_{\tau}$.
One of these originates from the Yukawa couplings
at the $H \tau \nu_{\tau}$ vertex. The second power
of $m_{\tau}$
originates from the trace over the lepton loop
resulting from the interference between the W
and the Higgs. This trace goes as:
$$
Tr[\gamma^\mu(\slp_\tau+m_\tau)(1-\gamma_5)\slp_\nu ]
= 4 m_\tau p_\nu^\mu
\eqnox\eqnA
$$
Clearly the $m_{\tau}$ in this trace arises
because one is summing over the spins of the final
$\tau$. So the $m_{\tau}$ in the above trace can be avoided
provided we do not sum over the spins of the $\tau$.
Thus we arrive at CP violating transverse polarization
asymmetries of the $\tau$ that are larger by $\approx
m_t/m_\tau \approx100$ compared to PIRA requiring
therefore $\approx 100^2$ fewer top pairs. Thus the transverse
polarization asymmetries may well be observable with $\approx
10^4$ $t \bar t$ pairs rather than $\approx 10^7$ needed for PIRA.
\par
We now briefly recapitulate some aspects of the Weinberg Model that are
relevant
to this work\refnum\tyeref.
The model consists of three Higgs doublets enabling it
to possess a CP violating Higgs sector without
the flavor changing neutral currents.
%The Weinberg model for CP violation is the minimal model for CP violation in
%the Higgs sector which does not contain flavour changing neutral
%currents\refnum\refone.
%In the standard model, CP violation occurs exclusively through
%the CKM matrix, this implies that certain CP violating
%quantities are naturally small. For example, it is well known that
%the electric dipole moment of fermions are naturally small\refnum\reftwo
%in the standard model.
%
%
%
%\par
%In addition it has been pointed  out in \sqref\gordy
%that the CP violating transverse $\mu$ polarization in $K_{3\mu}$ decays
%is naturally small in the standard model.
%This is due to the fact that to obtain such a polarization at tree level
%one must interfere the W-boson with a charged scalar.
%Likewise, transverse polarization of the
%lepton would be suppressed in
%the semi-leptonic decays of the $b$ or $t$-quark.
%
%
%
%
%\bigskip
%\hbox{\bf The Model}
%The model which we will be considering is a version of the
%Weinberg model put forward by \sqref\tyeref.
%We will largely be considering here phenomena associated with the
%charged higgs sector and so our results may be easily generalized to
%other models which give rise to such charged scalars.
%
%
%
%\par
%The basic structure of the Weinberg model consists of three higgs doublets
%which is chosen since it is the minimal case which can produce
%a CP violating higgs sector while protecting from flavour changing
%neutral currents.
It therefore has two charged Higgs states
$H_1^\pm$ and
$H_2^\pm$.
For simplicity we will assume that
$H_1^\pm$ has a much higher mass than $H_2^\pm$ ($m_H$)
so that we need to
consider effects due
to $H_2^\pm$ only.
%\par
The Lagrangian coupling $H_2$ to quarks and leptons is given\ref\tyeref
by:
$$
{\bf L }= {g\over\sqrt{2}} H_2^+ (
\bar u_i  {m_{dj} \over m_W} U_d P_R d_j V_{ij}^{KM}+
\bar u_i  {m_{ui} \over m_W} U_u P_L d_j V_{ij}^{KM}+
\bar\nu_i {m_{li} \over m_W} U_l P_R l_j\delta_{ij} )
+H.C.\eqnox\lagrangian
$$
where
$$
\eqalignno{
U_l=& -{c_1s_2s_3+c_2c_3e^{i\delta}\over s_1s_2}  \cr
U_u=& {c_1c_2s_3-s_2c_3e^{i\delta}\over s_1 c_2} \cr
U_d=& {s_1s_3\over c_1}
&(\xnum\eqnB)  \cr
}
$$
and $V^{KM}$ is the Kobayashi-Maskawa matrix.
We denote
$s_i=\sin(\theta_i)$ and
$c_i=\cos(\theta_i)$
where $\theta_i$ and $\delta$ are parameters of the Higgs potential.
The CP violation which we consider
is proportional to
combinations of these couplings such as
$
V_{ul}\equiv Im(U_u^*\ U_l).
$
\par
%
%))))))))))))))))))))))
%
%\hbox{\bf Top Decay }
%\par
%Consider now the case of the top decay
%$$
%t(p_t)\rightarrow
%\tau^+(P_\tau)\nu_\tau(p_\nu)b(p_b).
%$$
%
%
%
%
%DAVID, PLEASE CHECK language below. I thought tau momentum
%was in x direction, NOT top?
In the rest frame of the $\tau$ lepton, let us define a reference frame where
the momentum of the top quark is in the $-x$ direction, the $y$ direction is
defined to be in the decay plane such that the $y$ component of the $b$
momentum
is
positive and the $z$ axis is defined by the right hand rule.  In the limit that
the $\tau$ mass is small, (WM) can give rise to two kinds of
CP violating
polarization
asymmetries.
These are
\def\taua{\tau^+(\uparrow) \ }
\def\taub{\tau^+(\downarrow)\ }
\def\tauc{\tau^-(\uparrow) \ }
\def\taud{\tau^-(\downarrow)\ }
$$
\eqalignno{
A_{Y}&={\taua-\taub+\tauc-\taud\over\taua+\taub+\tauc+\taud}\cr
A_{Z}&={\taua-\taub-\tauc+\taud\over\taua+\taub+\tauc+\taud}
&(\xnum\eqnD)\cr
}
$$
where for $A_Y$ ($A_Z$) the arrows indicate the spin up or down
in the direction $y$ ($z$).
While both $A_Y$ and $A_Z$ are CP odd, being $T_N$ even, $A_Y$ needs an
interaction phase; $A_Z$ is odd under $T_N$ and does not need an
interaction phase\refnum\gordytwo. Thus $A_Y$($A_Z$) is proportional to
absorptive
(dispersive) part of Fig. 1a.
\par
For our analysis of top decay, we will ignore the masses of the b quark
and the $\tau$ lepton (wherever that is a good approximation). Thus let
us define the invariants:
$$
s=2p_\nu\cdot p_\tau\ \ \ \
t=2p_\tau\cdot p_b\ \ \ \
u=2p_b\cdot p_\nu
\eqnox\eqnE
$$
as well as the quantities
$y_W={\Gamma_W^2 / m_t^2}$,
$\lambda=s/m_t^2$ and
$x_i={m_i^2 /m_t^2}$
for $i=b$, $\tau$, $H$ and $W$.
The angle $\theta$ is defined to be
the angle between the $W$ momentum and the $\tau$
momentum in the $W$ rest frame.
\par
Consider now the case of the CP violating polarization in the $x-y$ plane.
This gives
rise to a polarization asymmetry in the direction
$$
V={sp_b-tp_\nu\over\sqrt{stu}}
\eqnox\eqnG
$$
which lies in the $x-y$ plane.
In the $\tau$ rest frame
if $\psi_b$ is the azimuthal angle of the b quark
momentum from the $x$ axis and $\psi_\nu$ is the angle of the $\nu$
then $V$ defined above  is a unit vector at angle
${1\over 2}(\psi_b+\psi_\nu-\pi)$
in the $x-y$ plane.
Note, however that since the $\tau$ rest frame is highly boosted with respect
to the top quark rest frame, the
b and $\nu$ are close to the $-x$ direction so $V$ will be very close to the
$y$ axis.
The differential asymmetry is thus given by (using here and in the relevant
formulas to follow $BR(W \rightarrow \tau\nu)=1/9$):
$$
dA_{Y}={9 g^2\over 32\pi^2}V_{ul}{(1-\lambda)^2\sqrt{x_\tau \lambda y_Wx_W}
\sin{\theta}\over
(1-x_W)^2(1+2x_W)(x_H-\lambda)((\lambda-x_W)^2+y_Wx_W)}
d\lambda\ d\cos\theta
\eqnox\eqnH
$$
Integrating this over $\theta$ and $\lambda$ using the narrow resonance
approximation the result for the total asymmetry is
$$
A_{Y}= {9\over 64}g^2 V_{ul} {\sqrt{x_\tau x_W}\over (1+2x_W)(x_H-x_W)}.
\eqnox\eqnI
$$
\par
The CP violating polarization asymmetry
in the $z$-direction
%arises from the interference of the same diagrams however
is proportional to the real part of the resonant $W$ propagator.
The differential asymmetry is:
$$
dA_{Z}=
-
{9 g^2\over 32\pi^2}V_{ul}
{ \sqrt{x_l}\sin\theta (\lambda-x_w)(1-\lambda)\sqrt{\lambda}
\over
(1-x_W)^2(1+2x_W)((\lambda-x_W)^2+x_Wy_W)(x_H-\lambda)}
d \lambda\ d\cos\theta   .
\eqnox\eqnJ
$$
Integrating this
over $\lambda$ and $\theta$ one obtains the total asymmetry
$$
A_{Z}=- {9\over 64\pi}g^2 V_{ul}
{\sqrt{x_l} \over (1-x_W)^2(1+2x_W)x_H}f   (x_W,y_W,x_H)
\eqnox\eqnK
$$
where $f$ is the integral
$$
f(x_W,y_W,x_H)
=\int_0^1 {(\lambda-x_W)x_H(1-\lambda)\sqrt{\lambda}
\over ((\lambda-x_W)^2+x_Wy_W)(x_H-\lambda)}d\lambda.
\eqnox\eqnL
$$
\par
Note that due to the dependence on the real part of the resonance propagator,
as $s$ passes through
$x_W$, the net polarization changes sign. There is therefore a partial
cancelation of
the polarization as defined above.
If however the invariant mass of the $\tau \nu$ system can experimentally be
determined within a few GeV, this information may be taken into account by
weighting events where $s\leq x_W$ with $-1$ and events where $s\geq x_W $
with $+1$. The total asymmetry defined in this way, $A_{Z}^{\prime}$, is thus
given by
$$
A_{Z}^{\prime}= -{9\over 64\pi}g^2 V_{ul}
{\sqrt{x_l} \over (1-x_W)^2(1+2x_W)x_H}f^\prime   (x_W,y_W,x_H)
\eqnox\eqnM
$$
where $f^\prime$ is the integral as equation ($\eqnL$) except
$\lambda-x_W$
is replaced by
$|\lambda-x_W|$.
Note that whereas $f$ is smooth as $y_W\rightarrow 0$, $f^\prime$ grows
logarithmically since in this case the resonant enhancement is not cancelled.
%
%
%$$
%f^\prime(x_W,y_W,x_H)
%=\int_0^1 {|\lambda-x_W| x_H(1-\lambda)\sqrt{\lambda}
%\over ((\lambda-x_W)^2+x_Wy_W)(x_H-\lambda)}d\lambda.
%\eqnox\eqnN
%$$
%
%
%
\par
Of course the polarization of the $\tau$ is not directly observable
and must be inferred from the decay distributions of the $\tau$.
Let us define the sensitivity $S$
of a method of measuring the polarization of the
$\tau$
so that given $N_\tau$
\relax $\tau$ leptons, the error in the
measurement of the polarization, $\Delta P$, is given by
$\Delta P= (S \sqrt N)^{-1}$.
In a study \refnum\taupol the decay modes
$\pi\nu$, $2\pi\nu$, $3\pi\nu$, $e\nu\bar\nu$ and
$\mu\nu\bar\nu$
are considered
as polarization analyzers.
The sensitivity which could be obtained
in an ideal detector is found to be
$S=0.25$ if one considers only the mode
$2\pi\nu$
while if one combines information from all four decay modes, $S=0.35$.
%
%     L3     Phys Lett B294 466 (1992)
%     ALEPH  Phys Lett B265 430 (1991)
%     A Rouge Workshop on Tau Lepton Physics, Orsay France, 1990
%
Thus, the error in measuring a polarization asymmetry $A$, given $N_t$ top
quarks, is given by
$$
\Delta A = (S \sqrt{N Br(t\rightarrow \tau\nu b)})^{-1}.
\eqnox\eqnO
$$
\par
In  [\pra] the
measurement of CP violation in this model is considered in
$t\rightarrow\tau\nu b$
without using
polarization information from the $\tau$.
CP violation is manifested by a difference in the distribution of $\tau$
leptons
in $\cos\theta$ leading to an asymmetry in the partially integrated
rate defined through the quantity $\alpha_+$ as

%A CP violating
%asymmetry which is therefore considered  is
%the quantity $\alpha_+$ is defined as
$$
\alpha_+ = {
 \Gamma_f(t\rightarrow b \tau^+ \nu_\tau)
-\Gamma_f(\bar t\rightarrow b \tau^- \nu_\tau)
\over
 \Gamma_f(t\rightarrow b \tau^+ \nu_\tau)
+\Gamma_f(\bar t\rightarrow b \tau^- \nu_\tau)
}
\eqnox\eqnP
$$
where $\Gamma_f$ is the rate
for the decay
into a state where
the angle between the $t$ momentum and the $\tau$ momentum in the $W$ rest
frame is greater than $\pi\over 2$.
The value of this asymmetry is:
$$
\alpha_+={9\sqrt{2}\over 4\pi}
{G_F m_\tau^2 r_{WH} V_{ul}\over (2+r_{Wt})(1-r_{WH})}
\eqnox\eqnS
$$
where $r_{WH}=m_W^2/m_H^2$ and $r_{Wt}=m_W^2/m_H^2$.
\par
All these quantities are proportional to $V_{ul}$. As in [\pra]
we now consider what the maximum value of $V_{ul}$ is
that is consistent with experimental
observations.
It turns out that the one constraint comes from
$\tau\rightarrow l \nu_l \nu_\tau$ decays where a 2\% violation of lepton
universality is not excluded by experiment\refnum\tauuni.
We therefore impose
$$
{\Gamma(\tau\rightarrow \mu \nu_\mu \nu_\tau)_{W+H}
-\Gamma(\tau\rightarrow \mu \nu_\mu \nu_\tau)_W
\over
\Gamma(\tau\rightarrow \mu \nu_\mu \nu_\tau)_W}\leq 0.02
\eqnox\eqnR
$$
where the subscript $H+W$ means the total rate including both $H$ and $W$
exchange while the subscript $W$ means
only the $W$ exchange (ie the SM). This gives rise to the condition
%%DAVID: Please check this inequality, something is definitely
%%wrong.
$$
{m_\mu^2 |U_u|^2|U_l|^2 \over M_H^2}
\left |
{m_\tau^2 |U_u|^2|U_l|^2\over 4 m_H^2} -2
\right | \leq 0.02.
\eqnox\eqnRR
$$
Another constraint comes from the non-observation of $b\rightarrow s\gamma$.
This decay in the WM model is calculated in \sqref\polish.
Following this paper we take $BR(b\rightarrow s\gamma)\leq 8.5\times10^{-4}$.
In addition perturbative limits on the the charged higgs fermion couplings
imply:
$$
|U_u|m_t,\ \
|U_d|m_b,\ \
|U_l|m_\tau \leq {4\pi m_W\over \sqrt{2} g}
\eqnox\eqnQ
$$
\par
The result thus obtained is $V_{ul}=10^3$. Using this value of $V_{ul}$
Table 1 shows (see the second column)
the value of each of the
asymmetries
$\alpha_+$, $A_{Y}$, $A_{Z}$ and $A_{Z}^{\prime}$
for
$m_t=150 GeV$ and $m_H=200$ and $300GeV$.
We then show how many top
quarks are needed to observe these asymmetry where we take the sensitivity to
$\tau$ polarization $S=0.25$ which is obtained using only the $2\pi\nu$ mode.
We also calculate the restriction which may be placed on $V_{ul}$ using $10^6$
top quarks (see the last column). We see that
the polarization asymmetries place much
more stringent restrictions than $\alpha_+$
being one to two orders of magnitude more effective.

%
%
%
%
%
%      b-stuff
%
%
%
%
%
\par
%CP-violating signals of the weinberg model have previously\refnum\dono
%been considered in the context of B-physics. In [\dono] CP violation is
%produced by a gluon penguin graph and results in partial rate asymmetries in
%hadronic decay modes. In this case the results are proportional to $V_{du}$.
CP violating
$\tau$-lepton polarization asymmetries may also be considered
in
semileptonic decays of B mesons to $\tau$ via:
$B \rightarrow \tau + \nu + X$.
Now the signal will be largely proportional to $V_{ld}$\refnum\dono.
Since for semileptonic B decays the spectator model
works quite well it is sufficient for our purpose to
study the free quark decay,
$b\rightarrow\tau\nu c$.
In this case,
the $W$ is far off shell so that the asymmetries
which depend on interaction phases are higher order in the coupling compared to
asymmetries which do not require such a phase.
Thus, polarization asymmetries in
the $x-y$ plane and energy asymmetries of the $\tau$ lepton
are expected to be
much smaller than the polarization asymmetry in the $z$ direction.
The value of this asymmetry may be readily calculated as in the case of the
top quark giving
$$
A_{Z}={8\pi\over 35} (V_{ld}-{m_c^2\over m_b^2}V_{ul})
{\sqrt{u_\tau}\over u_H}
{J(u_\tau,u_c)\over I(u_\tau,u_c)}\eqnox\beqA .
$$
where $u_i=m_i^2/m_b^2$ for $i=\tau$, $c$ and $H$,
$I$ and $J$ are the kinematic integrals:
$$
\eqalignno{
I(u_c,u_\tau)&=
12\int_{r_0}^1 (r-u_c-u_\tau)(1-r)^2\lambda^{1\over 2} (r,u_\tau,u_c)
{dr\over r}
\cr
J(u_c,u_\tau)&=
{105\over 16}\int_{r_0}^1 r^{-{3\over 2}}(1-r)^2\lambda(r,u_\tau,u_c)\ dr
&(\xnum\beqC)\cr
}
$$
Here
$$
r_0=(u_c^{1\over 2}+u_\tau^{1\over 2})^2;\ \ \ \
\lambda(a,b,c)=a^2+b^2+c^2-2ab-2bc-2ca.
\eqnox\beqD
$$
%
%
%
%The branching ratio for $b\rightarrow \tau\nu c$
%is given by (assuming only the standard model contribution):
%$$
%Br(b\rightarrow\tau\nu c)={I(u_c,u_\tau)\over I(u_c,0)}
%Br(b\rightarrow   e\nu c)\approx .046
%\eqnox\beqB
%$$
%
%
%
Evaluating the above expression we find that $A_{Z}=1.2\times 10^{-4} V_{ld}$
for $m_H=200 GeV$ and $A_{Z}=5.2\times 10^{-5}
(V_{ld}-{m_c^2\over m_b^2}V_{ul}) $
for $m_H=300 GeV$.  We thus note that asymmetries from $b$ decay will
primarily
put
restrictions on the quantity $V_{ld}$ because the factor
$m_c^2\over m_b^2$ suppresses the contribution of the term with $V_{ul}$
by about an order of magnitude.
Returning to the discussion of the existing experimental bounds on the
Weinberg model, we find that there is little restriction on
$V_{lb}$ from existing experiments; in fact it may be as large
as $1.3\times 10^4$.
Even if $V_{lb}$ is an order of magnitude less it would require
about $10^5$ B's (assuming the same efficiency of 0.25 for
detecting the $\tau$ polarization) to show up as a non-vanishing
effect.
%of $10^6$ B's  useful constraints can be put on this class of models.
At a B-factory,
given $10^8$ $b$ quarks, under ideal conditions (using just the $2\pi\nu$
decay mode to determine the polarization of the $\tau$), at $1\sigma$ one may
put a restriction of $V_{ld}\leq 16$ if $m_H=200GeV$ and $V_{ld}\leq 37$ for
$m_H=300GeV$.
We thus see that it takes about
100 times more $B's$ than top quarks
to attain a similar reach on the CP parameters, i.e. $V's$, of the extended
Higgs model.
\bigskip\bigskip
\noindent{\bf Acknowledgement}:
This research
was supported in part by the U.S.-Israel
Binational Science Foundation.
The work of D.A. is supported in part by an
SSC Fellowship and USDOE contract DE-AC-76SF00515.
The work of D.A.\ and A.S.\ is
supported in part by USDOE contract number DE-AC02-76CH0016. The work
of G.E.\ is supported in part by the Smoler Research Fund and by the
Fund for the Promotion of Research at the Technion.
%
%%
%%
%
%%%%%%%%%%%%%%%%%%%%%%%%%%%%%%%%%%%%%%%%%%%%%%%%%%%%%%%%%%%
\vfill\eject
\centerline{\bf Table 1}
\bigskip
\bigskip
Here the value for the asymmetries (second column)
above are the maximum consistent with the
restrictions described above (i.e. using $V_{ul}=10^3$).
$N_t$ is the number of top quarks required to see the
asymmetry, given in the second column,
taking into account the sensitivity to $\tau$ polarization using only
the $2\pi\nu$ decay mode. $V_{ul}^{max}$ is the restriction on $V_{ul}$
that may be attainable with $10^6$ top quarks.
In all cases $m_t=150GeV$, the upper number in each case is for $m_H=200GeV$
and the lower is for $m_H=300GeV$.
\bigskip
\bigskip
\bigskip
\bigskip
\settabs 4 \columns
\+ Asymmetry & Value($10^{-3}$) & $N_t$ & $V_{ul}^{max}$ \cr
\medskip
\+ $\alpha_+$ &  $2.8$ & $3.2\times10^6$ & $1800$ \cr
\+            &  $1.4$ & $13 \times10^6$ & $3600$ \cr
\medskip
\+ $A_{Y}   $ &  $161$ & $5.6\times10^3$ & $ 75  $\cr
\+            &  $ 65$ & $3.4\times10^4$ & $184  $\cr
\medskip
\+ $A_{Z}   $ &  $ 96$ & $1.6\times10^4$ & $126  $\cr
\+            &  $ 29$ & $1.7\times10^5$ & $412  $\cr
\medskip
\+ $A_{Z}^{\prime}   $ &  $540$ & $4.9\times10^2$ & $ 22  $\cr
\+                    &  $220$ & $3.0\times10^3$  & $ 55  $\cr
\vfill
\eject
\centerline{\bf References}
\bigskip
\item{\ehs.} G. Eilam, J. Hewett and A. Soni, Phys. Rev. Lett. 67, 1979 (1991)
and 68, 2103(1992) (Comment).
\medskip
\item{\km.}M. Kobayashi and T. Maskawa, Prog. Theor. Phys. 49, 652(1973).
\medskip
\item{\nelson.} See e.g. A. E. Nelson, D. B. Kaplan and A. G. Cohen,
Nucl. Phys. B373, 453 (1992) and references therein.
\medskip
\item{\aas.} D. Atwood, A. Aeppli and A. Soni, Phys. Rev. Lett. 69, 2754
(1992).
\medskip
\item{\pra.}
D. Atwood, G. Eilam, A. Soni, R. Mendel and R. Migneron, BNL-48162 (1992),
to appear in Phys. Rev. Lett.
\medskip
\item{\jg.} R. Cruz, B. Grzadkowski and J. F. Gunion, Phys. Lett. B
289, 440(1992).
\medskip
\item{\others.} G.~Kane, G.~Ladinsky and C.P.~Yuan, Phys. Rev.
D45, 124 (1991); C.~Schmidt and M.~Peskin, Phys. Rev. Lett.
69, 410 (1992); B.~Grzadkowski and J.F.~Gunion, Phys. Lett.
287, 237 (1992);
W.~Bernreuther, O.~Nachtman, P.~Overmann and T.~Schr\"oder,
preprint HD-THEO-92-14; N.G.~Deshpande, B.~Margolis and H.D.~Trottier,
Phys. Rev. D45, 178 (1992); C.R.~Schmidt, SLAC-PUB-5878.
\medskip
\item{\refone.} S. Weinberg, Phys. Rev. Lett. 37, 657(1976); see also
            T. D. Lee, Phys. Rev. D8, 1226 (1973).
\medskip
\item{\rxu.} A. Soni and R. M. Xu, Phys. Rev. Lett. 69, 33(1992).
\medskip
\item{\mpcs.} See also M. Peskin and C. Schmidt, ibid.
\medskip
\item{\mhgemt.} For simplicity we are assuming $m_H > m_t$,
thereby Fig. 1b is purely dispersive.
\medskip
\item{\seepra.} See Atwood et. al.
Ref. [\pra]
in particular eqns. 11 and 12.
\medskip
\item{\tyeref.} C. H. Albright, J. Smith and S-H. H. Tye, Phys. Rev. D21, 711
(1980).
\medskip
\item{\gordytwo.} Note that a polarization asymmetry similar
to $A_Z$ was considered in the context of $K_{\mu3}$ decays
in R. Garisto and G. Kane, Phys. Rev. D44, 2038(1991).
\medskip
\item{\taupol.} A. Rouge, Workshop on Tau Lepton Physics, Orsay France (1990)
p 213. Note also that tau polarizations have actually been measured
at LEP see
ALEPH Collaboration, Phys. Lett. B265, 430 (1991) and L3 Collaboration
Phys. Lett. B294, 466 (1992).
\medskip
\item{\tauuni.} W. J. Marciano, Phys. Rev. D45, R721 (1992).
\medskip
\item{\polish.} P. Krawczyk and S. Pokorski, Nuc. Phys. B364, 10 (1991).
\medskip
\item{\dono.}
CP violation in B-decays in the Weinberg model has previously
been considered by
J. F. Donoghue and E. Golowich, Phys. Rev. D37, 2542 (1988).
However, their effects are driven by $V_{du}$ so our work
complements theirs.
\vfill\eject
\hbox{\bf Figure Caption}
\bigskip
Figure 1a: The Feynman diagram for $t\rightarrow b \tau^+ \nu_\tau$ through
$W^+$ exchange (i.e. the SM mechanism).
Figure 1b: The Feynman diagram for $t\rightarrow b \tau^+ \nu_\tau$ through
$H^+$ exchange. For b decays via $b \rightarrow c \tau \nu$,
replace $t \rightarrow b$ with $b \rightarrow c$ in the two diagrams.
\vfill\eject
%
%
%
%
%
%
%
%  postscript version of figure follows \end command
%
%
%
%
%
\end